# Trends of development of the methodical system of teaching physics in high schools of Kazakhstan

*KerimbayevNurassylNurymuly*
*Kazakhstan, Astana*

Abstract
  The research study in this article is about the development trends of the methodical system in teaching Physics in high schools of Kazakhstan, and taking into account the educational process informatization. This article shows the value of methodical system in teaching Physics in high school as one factor of the students' erudition level increasing, and the professional competence formation.

Modern Education of the Republic of Kazakhstan in terms of socio-economic and information technology transformation requires increasing methodological and system knowledge. One of the major changes of physical education should be the methodological orientation. Physics theory and method of teaching should be directed to the increasing complexity of scientific knowledge, the changing balance between research and presentation of scientific knowledge.

The Fundamentals of physical thinking are formed and developed in specific forms and methods of training. Methodical system of teaching of physics in high school is based on a scientific basis, considering the subject matter of a factual, ideological and methodological point of view. The development of technical training system of high school physics creates the necessary prerequisites for raising the educational level of students, the formation of professional competence.

Scientific and methodological problems of informatization process of teaching physics in high school is the problem of the formation and development of basic and professional skills of future professionals. The process requires scientific information, training and methodological support.

The introductions of information and communication tools are changing organizational forms of the learning process.

The Trends in development of methodical physicsteaching of in high schools of Republic of Kazakhstan are in the following positions:
• the process of physical education is based on conceptual and methodological research;
• methodical system of teaching physics should be based on theoretical and information modeling;
• learning theory of physics developed through the use of computer technology.

Physics course in modern higher education of Kazakhstan includes effective teaching and learning information technology, focused on the formation of key competencies, which in turn develop the motivation, proactive learners.

New information technology teaching physics that are the subject of information, promote processes such as design, construction, implementation, analysis, rendering computer systems development methodology of teaching physics at the university.

The learning process in higher education should be based on this concept and methodology of training that combines theory of design methodology and theory of mastering practical skills. In the learning process of information solves a number of methodological problems and is performed based on educational technology, using of information-communication technologies and allow toorganize collective and individual students. Active involvement of students in activities to address the challenges of innovation extends content-semantic space of the learning process.

The goal of physical education is to develop students' knowledge of the theoretical basis of the methods of teaching physics. The content and organization of the educational process in physics in the framework of modern educational technology is the training of specialists in teaching physics in the modern school. The study of theoretical questions accompanied practical training, which worked through the implementation of all kinds of school physics experiment.

The components of the methical system of teaching physics at the university are learning objectives, content of physical education, methods, tools and organizational forms of education. Methical system for the preparation of future teachers of physics based on scientific and theoretical positions of the discipline (content of the training material, exercises, laboratory work and demonstration). Listed components methical training system aimed at creating motivation and incentives for students to acquire knowledge independently. Methical system aims to develop skills such as the ability to design a lesson, teaching technology and teaching methods, teaching and educational activities and other kinds and forms of academic and extracurricular activities.

The choice of main components of methodical system of training future teachers of physics are relying on modern educational technologies, including information and communication.

Methodical development of physical education in the Republic of Kazakhstan, in our opinion, has the following areas:

- studying the scientific-theoretical and methodological bases of information for physical education;
- theoretical modeling of methodical system of teaching physics;
- technological approach and trends in methods of teaching physics in the educational environment;
- involvement in the search, research and self-research students using new technologies;
- Developing training methods solving physics problems with a computer and the use of computer simulation training methodology;
- Use of the computer in physics teaching and research activities of students.

Teaching physics at the selection of material is based on the need to introduce students to the field of "Physics and mathematics education" with modern scientific data. Planning and training sessions are held on the physics-specific topics and sections of the educational program and in accordance with the curriculum, the use of technical training, new teaching, information and computer technology.

Methical system of teaching of physics should include different types of control and the use of modern means of evaluating the results of physics teaching, the organization of control over the results of training and education.

Educational Technology and IT act as a bridge for the purpose of education of future teachers of physics. Learning technologies include forms of differential and individual learning, especially the teaching of physics in classes of different profiles.

Students, future teachers of physics, should know the basics of school education and preprofile training, elective courses of different directions, content, design features of programs, methods of teaching and elective courses.

System methodical preparation of future teachers of physics is aimed at:
• to help students the ability to set educational goals and objectives;
• the ability to analyze the modern teaching kits in physics for secondary school, exercising choice in terms of their compliance with the objectives of teaching physics, didactic principles and age of students;
• ability to select and design the technology and methods of teaching, designing various models of physics lessons.

Future physics teacher should also be able to plan educational work, to increase the cognitive activity of students in the classroom, to be able to choose differentiated homework and organize extracurricular activities of students in physics.

Improving the content of methodological training of future physics teachers in the pedagogical university should be by strengthening the theoretical and methodological elements. Methical preparation involves selection of invariant and universal in its essence the fundamental methodological knowledge and skills, determination of system concepts, knowledge and skills of school physics course. Methical system of training high school physics involves mastering a generalized way of professional and methodological activities, providing the solution set of specific problems of the domain.

Information content of discipline due to the general objectives of bachelor degree in "Physics", the requirements of the state educational standard of higher education in the appropriate direction, content and methodology of modern methods of teaching physics as an integrated science.

Level of development of fundamental knowledge and teaching skills, professional and personal qualities will enable future teachers of physics to solve creative problems methodical nature and effectively implement them in a professional and educational activities.

The quality and effectiveness of professional and educational activities also depends on the level of formation of key competence. Today education content requires system knowledge, skills and methods of their implementation. Building Committee-activity due to personal and individual approach to education. The presence of key competencies of future teachers is defined as a learning activity, and

in a specific teaching situations, actions and, of course, fully manifested in direct teaching activities.[2]

Competence is characterized by willingness and ability to address new, practically important tasks. Competence requires the ability to evaluate the results of their activities, refleksirovanie level in their area of competence. The concept of "core competence" of its components include the following competencies:
• information;
• communicative;
• engineeringdesign;
• estimated.

We emphasize that the formation of these skills is one of the ways of methodical system of teaching physics. In this paper, we consider the formation of key competences in physics by using information technology.

Modern education requires the use of active-activity methods, the use of problem approaches, training, taking into account the features of the individual student, which means that the traditional class-task system with its methods is now ineffective and insufficient.

There is one solution to this problem is the use of information and communication technologies, organizing creative, educational and research activities of students, with a number of advantages over paper and other technical means of education, namely:

1) The material presented in the form of multimedia, providing a high level of visualization, the possibility of varying time scales of events, interrupts an interactive models, etc.
2) Analysis, testing and correction of results.
3) The use of software environments, virtual laboratories. Modeling makes up the shortage of equipment and reagents, is safe and is indispensable in the study of micro and macro world, the processes occurring at different rates.
4) Login to the training system by their password and login you can create individual learning paths for each student, that is, implement individualized learning.
5) Communication via network connects students with the teacher, outside consultants, remote equipment.

Thus, the development of key components of the methodological training of future teachers of physics should be given information of the learning process, the introduction of information and communication tools, training, changing the organizational form of learning and contributing to the development of future teachers of physics.